# Interpretable Link Prediction in AI-Driven Cancer Research: Uncovering Co-Authorship Patterns

Shahab Mosallaie[1], Andrea Schiffauerova[1] and Ashkan Ebadi[1,2,*]

[1] *CIISE, Concordia University, Canada*
[2] *Digital Technologies, National Research Council Canada, Canada.*

**\*** Email: ashkan.ebadi@nrc-cnrc.gc.ca

**Abstract**

Artificial intelligence (AI) is transforming cancer diagnosis and treatment. The intricate nature of this disease necessitates the collaboration of diverse stakeholders with varied expertise to ensure the effectiveness of cancer research. Despite its importance, forming effective interdisciplinary research teams remains challenging. Understanding and predicting collaboration patterns can help researchers, organizations, and policymakers optimize resources and foster impactful research. We examined co-authorship networks as a proxy for collaboration within AI-driven cancer research. Using 7,738 publications (2000-2017) from Scopus, we constructed 36 overlapping co-authorship networks representing new, persistent, and discontinued collaborations. We engineered both attribute-based and structure-based features and built four machine learning classifiers. Model interpretability was performed using Shapley Additive Explanations (SHAP). Random forest achieved the highest recall for all three types of examined collaborations. The discipline similarity score emerged as a crucial factor, positively affecting new and persistent patterns while negatively impacting discontinued collaborations. Additionally, high productivity and seniority were positively associated with discontinued links. Our findings can guide the formation of effective research teams, enhance interdisciplinary cooperation, and inform strategic policy decisions.

## 1. Introduction

Cancer remains a leading cause of death globally, accounting for nearly 10 million fatalities in 2020 (Ferlay et al. 2021), representing approximately one in every six deaths recorded worldwide (World Health Organization 2022). In Canada, cancer continues to hold its position as the primary cause of death, with approximately 40% of Canadians expected to receive a cancer diagnosis in their lifetime, and about 25% likely to succumb to the disease (Government of Canada 2021). Despite its lethal nature, early detection and effective treatment can lead to the successful cure of many patients and cancer types (World Health Organization 2022). In fact, timely diagnosis and intervention can arrest cancer progression and enhance prognosis (Crosby et al. 2022). This highlights the urgent necessity for effective cancer diagnostic strategies and procedures to save countless lives from this life-threatening disease (Okoli et al. 2021).

The rapid evolution of advanced cancer treatment technologies further underscores the importance of interdisciplinary collaboration. For example, nanoparticle-based drug delivery systems have emerged as a promising approach to improve therapeutic precision and reduce systemic toxicity in oncology. These systems can be engineered to target specific tumour sites, enabling controlled release of therapeutic agents and potentially improving patient outcomes (Cheng et al. 2021). The development of such technologies requires close collaboration among oncologists, materials scientists, biomedical engineers, and data scientists, particularly in the design, testing, and optimization phases.

Enhanced health outcomes can be achieved through collaboration among healthcare professionals (Ebadi et al. 2017), facilitated by knowledge sharing and improved decision-making. Cancer care, being a complex and fragmented process (Ullgren 2021), requires healthcare providers and patients to work together as a cohesive team to deliver high-quality care (Amafah et al. 2023). Furthermore, effective cancer care, encompassing both diagnosis and treatment, requires collaboration among researchers and professionals from diverse disciplines, each contributing their unique expertise (Knoop, Wujcik, and Wujcik 2017). Many AI-driven healthcare research projects are inherently international in scope, bringing together multi-country teams that integrate diverse expertise, resources, and perspectives to address complex scientific and clinical challenges.

Considerable effort has been dedicated to understanding the dynamics of research collaboration and pinpointing the factors that encourage it. Several studies credited innovation and scientific discoveries to the collaborative efforts of individual researchers (e.g., Beck et al. 2022). Studies also highlighted the positive impact of collaboration on research productivity (e.g., Ebadi and Schiffauerova 2016). Co-authorship is frequently used as a proxy for scientific collaboration (Ebadi and Schiffauerova 2015; Essers, Grigoli, and Pugacheva 2022), a trend further facilitated by the availability of large-scale bibliographic data. Research into the determinants of co-authorship typically focuses on two perspectives: one based on social dynamics (e.g., Lalli, Howey, and Wintergrün 2020) and the other derived from co-authorship networks (e.g., Ebadi and Schiffauerova 2015). These perspectives are interconnected, suggesting that co-authorship networks may emerge from underlying social dynamics.

At the macro level, collaborative research, whether or not it results in co-authorship, is driven by three primary factors: intellectual, economic, and social (Thakur, Wang, and Cozzens 2011). Intellectual drivers are often characterized by specialization and multidisciplinary collaboration (Thakur et al. 2011). Economic factors, such as substantial funding, can promote increased collaboration and resource sharing (Davies et al. 2022). Social interactions are vital for both knowledge acquisition and career advancement. For instance, Thakur et al. (2011) highlighted the impact of social factors by illustrating how junior researchers collaborate with senior colleagues to access resources, e.g., equipment and data, as well as to gain prestige through partnerships with esteemed scientists. Additionally, other studies have shown that highly productive researchers tend to collaborate with peers who exhibit similar levels of productivity (Moed, Glänzel, and Schmoch 2005).

While Yu et al. (2014) explored the prediction of co-authorship links within the medical field, there remains a notable gap in research specifically targeting the co-authorship network of researchers focused on AI applications in cancer diagnosis and treatment. This area of study is crucial, given the complexity of cancer and the necessity for collaboration among researchers from various disciplines. Our study seeks to uncover the factors influencing different co-authorship patterns among researchers engaged in AI-driven cancer research. We analyzed three specific patterns:

- **New co-authorship pattern:** Establishment of collaboration links between authors who have not previously co-authored a publication.

- **Persistent co-authorship pattern:** Continued collaboration links between authors who have co-authored one or more papers in the past.

- **Discontinued co-authorship pattern:** Termination of collaboration links between authors who have previously co-authored one or more papers.

The primary objective of this study is to develop interpretable machine learning models for predicting three distinct co-authorship patterns, i.e., new, persistent, and discontinued collaborations, in AI-driven cancer research. In addition, the study aims to identify and interpret the key driving factors behind each co-authorship pattern using explainable AI techniques, providing actionable insights for researchers, institutions, and policymakers. This study is pivotal in enhancing our understanding of the dynamics within co-authorship networks, particularly in the interdisciplinary field of AI-driven cancer research. Such understanding is crucial for tackling the complexities of cancer diagnosis and treatment. By identifying and interpreting the factors influencing different co-authorship patterns, we provide valuable insights for forming effective research collaborations, optimizing resource allocation, and fostering innovation across diverse scientific fields. This study makes the following contributions: (1) we propose a methodological framework that integrates both attribute-based and structure-based features for predicting multiple types of co-authorship patterns, (2) by applying interpretability techniques, we enable the identification and ranking of features that most strongly influence each co-authorship pattern, (3) instead of focusing primarily on predicting new collaborations, our approach simultaneously models new, persistent, and discontinued collaborations, offering a more complete view of collaboration dynamics, and (4) we apply our framework to a large-scale dataset of publications in AI-driven cancer research, a domain characterized by high interdisciplinarity and significant societal impact.

The remainder of the paper is structured as follows: Section 2 provides an overview of the related works. Section 3 outlines the data pipeline in detail. The methodology is presented in Section 4. In Section 5, the findings are presented and discussed, showcasing the analysis of new, persistent, and discontinued co-authorship patterns with supporting data. Section 6 concludes the paper. Limitations of the study and directions for future investigations are presented in Section 7.

## 2. Related works

The literature employed various methods, such as interviews, surveys, and social network analysis (SNA), to identify factors influencing co-authorship. In co-authorship networks, nodes represent authors, and a link between two authors indicates a joint publication. These networks are dynamic, constantly evolving with the addition of new authors and links, while also experiencing potential cessation of collaborations. Link prediction involves estimating the likelihood of future connections between nodes in a network based on existing connections and network characteristics (Hasin and Hassan 2022). In co-authorship networks, researchers who share similarities are more likely to collaborate in the future. Predicted co-authorship links can serve as a guide for forming robust research teams, providing strategic directions for collaboration (Pavlov and Ichise 2007).

In another study (Liben-Nowell and Kleinberg 2003), similarity-based metrics were employed to predict future co-authorship links by analyzing five co-authorship networks from arXiv, covering the years 1994 to 1999. The findings revealed that most similarity-based metrics outperformed various baselines in predicting co-authorship links. Pavlov and Ichise (Pavlov and Ichise 2007) introduced an improved link prediction method by integrating multiple similarity-based metrics into a feature vector and utilizing supervised machine learning (ML) to forecast new co-authorship links. Their research focused on the co-authorship network of Japanese researchers from 1993 to 2006. Hasan et al. (2006) further advanced this approach by incorporating attribute-based similarity metrics into the feature vector, aiming to boost ML model performance in co-authorship link prediction. For instance, they considered the size of the intersection set derived from a pair of authors' publication keywords, concluding that authors with higher values are more similar, thus increasing the likelihood of future collaborations. In another study, Yu et al. (2014) extracted various similarity-based metrics from co-authorship networks in coronary artery disease research, using these features as input for two distinct ML models. They demonstrated that features based on network structure could effectively predict potential future collaborations.

In a study by Chuan et al. (2018), various similarity-based link prediction metrics were assessed, and a novel hybrid attribute-based feature was introduced to enhance co-authorship link predictions. This innovative feature incorporates the content of authors' publications by employing a Latent Dirichlet Allocation (LDA) topic model (Blei 2003) to extract publication topics and calculate a discipline similarity score for author pairs. The study indicated that researchers are more inclined to collaborate with peers conducting research in similar fields. By analyzing three co-authorship networks across different domains of physics, the study highlighted the importance of evaluating publication content to determine the similarity scores of authors and papers, thereby improving the accuracy of co-authorship link prediction.

Several studies have employed link prediction and ML techniques to forecast co-authorship links, primarily focusing on predicting collaborations among researchers who have not previously partnered. However, neglecting to consider the continuation or termination of collaborations among authors with a history of co-authoring may limit our understanding of the factors driving various co-authorship patterns. For predictive models to be effectively utilized by recommender systems, they should provide a comprehensive overview of all potential co-authorship links, facilitating the formation of strong research teams. Additionally, the opacity surrounding how ML classifiers make these predictions leaves the key metrics or factors influencing co-authorship link prediction unidentified. Interpretability is essential, particularly when these models are used in decision-making processes, such as strategic planning, identifying potential collaborators by researchers, and building cohesive research teams within organizations.

## 3. Data

To examine the co-authorship patterns among researchers involved in AI-driven cancer research, data were sourced from Elsevier's Scopus, encompassing publications from 2000 to 2017. This dataset comprises journal articles, conference papers, book chapters, and books, resulting in a total of 7,738 publications, 46,644 authors, and 123,054 co-authorship links. As illustrated in Figure 1, there is a noticeable upward trajectory in the number of publications, which has been increasingly accelerating in recent years.

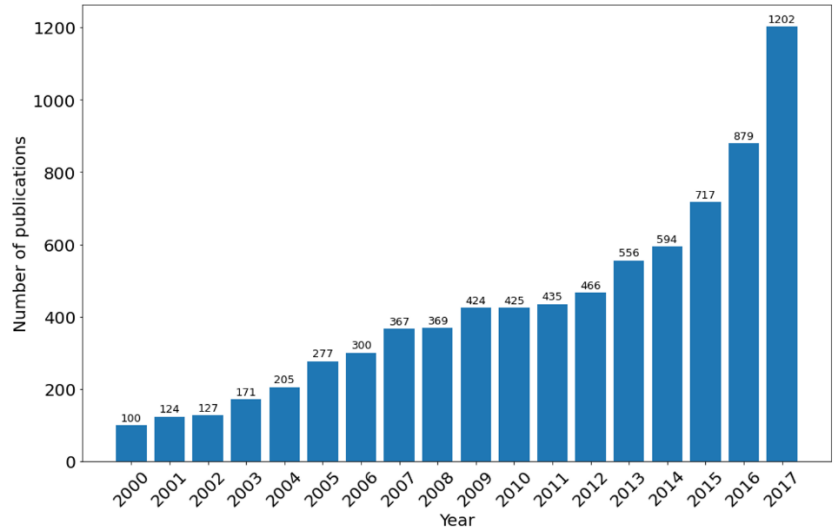

**Figure 1.** Publications over time.

## 4. Methodology

The data analytics workflow consists of four key components. First, the publication data is preprocessed to prepare it for analysis. Second, various features are engineered across different types. The third component involves developing ML models to predict co-authorship links. Finally, the fourth component is dedicated to interpreting these models to identify the most significant predictive factors. Figure 2 illustrates the high-level conceptual flow of the analyses. Each component is explained in detail in the following sections.

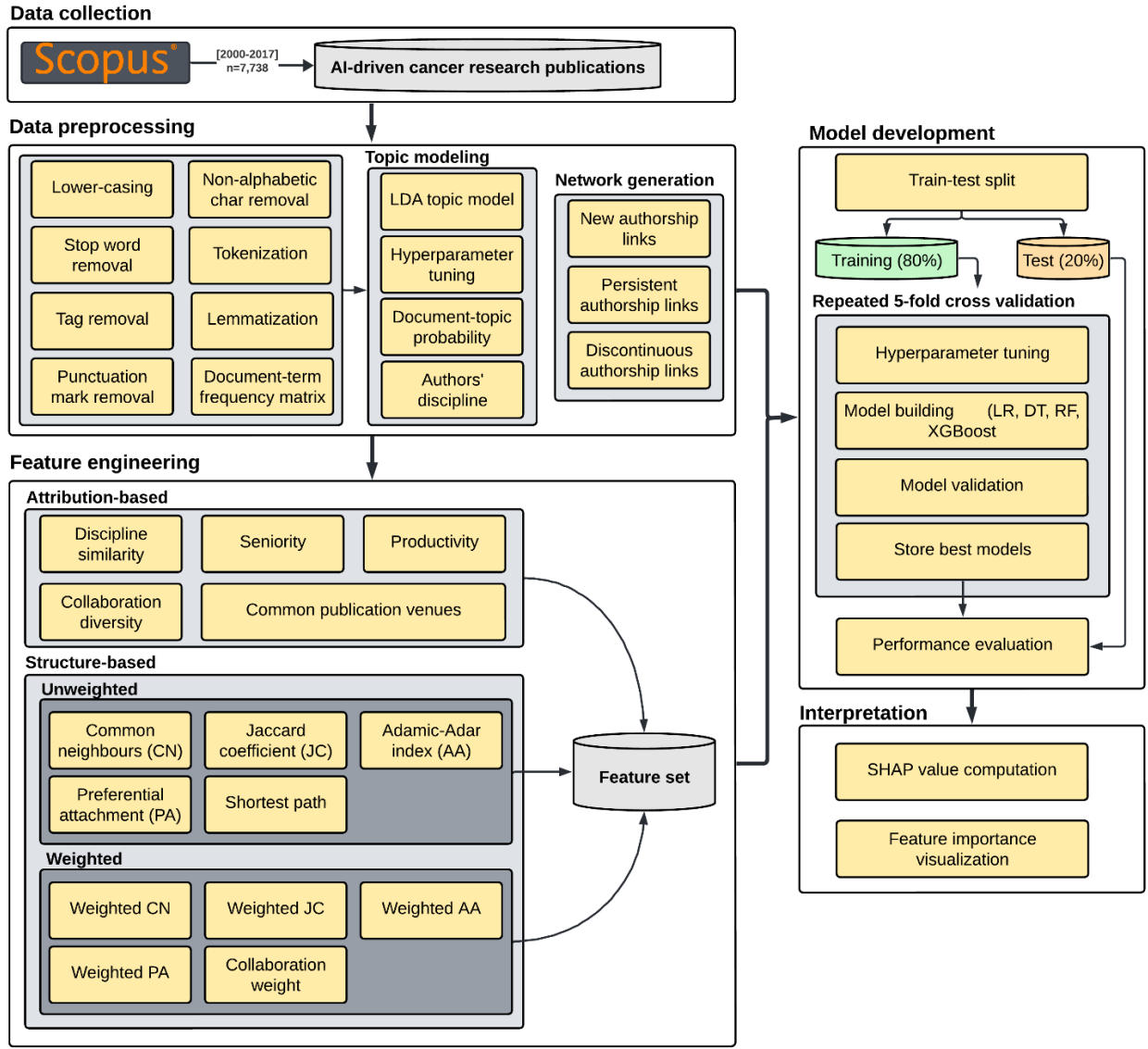

**Figure 2.** High-level conceptual flow of the analyses. The workflow includes four main components: 1) Data preprocessing, 2) Feature engineering, 3) Model development, and 4) Interpretation.

**4.1. Data preprocessing**

We applied several preprocessing steps to the textual data (i.e., titles and abstracts of publications), including converting text to lowercase and removing special characters, stop words, tags, and punctuation marks. Following this, the textual data was tokenized and lemmatized, resulting in a document-term frequency matrix.

In the data preparation phase, we constructed a total of 36 overlapping co-authorship networks, with 12 networks dedicated to each of the three co-authorship patterns: *new*, *persistent*, and *discontinued*. This was achieved using a methodology akin to that described in (Chuan et al. 2018; Pavlov and Ichise 2007). Each network utilized a four-year input window to extract features from past co-authorship activities and a three-year prediction window to assign labels that indicate whether co-authorship links were new, ongoing, or discontinued. This approach ensured the data was adequately prepared for machine learning models, with each dataset containing a feature vector composed of attribute-based and structure-based metrics, along with a label identifying specific co-authorship links. Our focus was on authors who were active during both the input and prediction windows, providing a comprehensive perspective of co-authorship dynamics and ensuring a well-labelled dataset.

For the new co-authorship pattern, labels were assigned such that class 1 indicates the future emergence of a new co-authorship link, while class 0 indicates its absence. A common challenge in predicting links within co-authorship networks is class imbalance, where one class significantly outweighs the other— for example, more author pairs not collaborating compared to those who do during the prediction windows. To tackle this imbalance, we excluded links between author pairs with no connecting path in their networks and employed random undersampling to balance the dataset, having tested various techniques. For the persistent and discontinued co-authorship patterns, we constructed 12 networks each. In the persistent pattern, class 1 signifies continued collaboration, while class 0 denotes its cessation. Conversely, in the discontinued pattern, class 1 marks the end of collaboration, whereas class 0 indicates its continuation, representing the opposite scenario of the persistent pattern.

**4.2. Feature engineering**

We extracted two groups of features from the preprocessed data: (1) Attribute-based, and (2) Structure-based.

*4.2.1. Attribution-based features*

These features focus on individual author characteristics, covering aspects such as discipline, seniority, productivity, collaboration diversity, and publication venues. Publication venues pertain to the specific conferences and journals where authors have disseminated their work. This feature set includes: discipline similarity score, seniority similarity score, seniority level of authors, productivity similarity score, productivity level of authors, collaboration diversity similarity score, collaboration diversity level of authors, and the number of shared publication venues. Each feature is elaborated upon in detail in this section.

**Discipline similarity:** Utilizing the document-term frequency matrix (as detailed in Section 4.1), we developed an LDA model (Blei 2003) and performed hyperparameter tuning, identifying six as the optimal number of topics based on coherence scores (Röder, Both, and Hinneburg 2015) and expert validation. The model produced a document-topic probability dataset, allowing us to determine each author's discipline by averaging the feature vectors of their publications. Subsequently, we calculated discipline similarity scores for author pairs using cosine similarity to assess the alignment of their research interests, grounded in the notion that authors are more inclined to collaborate with peers in similar fields. The discipline similarity score ranges from 0 to 1, where higher values denote a greater resemblance between the disciplines of an author pair.

**Seniority:** We utilized an author's career age as an indicator of their seniority level, determined by the duration between their first and most recent publications. From this, we developed two metrics for author pairs: the *seniority similarity score* and the *seniority level*. The seniority similarity score evaluates how closely matched a pair of authors in terms of career age, calculated by the absolute difference between their career ages, with lower values signifying greater similarity in seniority. In contrast, the seniority level metric reflects the combined seniority of an author pair, where higher values denote greater overall seniority. This metric is calculated as the sum of the career ages of both authors in the pair.

**Productivity:** An author's number of publications served as a measure of their productivity. From individual productivity levels, we derived two metrics for author pairs: the *productivity similarity score* and the *productivity level*. The productivity similarity score assesses how closely aligned a pair of authors is in terms of productivity, calculated as the absolute difference between their publication counts—lower values indicate greater similarity. Conversely, the productivity level metric reflects the combined productivity of the author pair, with higher values indicating greater overall productivity. This metric is calculated as the sum of their publication counts.

**Collaboration diversity:** Degree centrality served as an indicator of an author's collaboration diversity level, reflecting the number of distinct collaborations an author has participated in. Based on this, we developed two features: the *collaboration diversity similarity score* and the *collaboration diversity level*. The collaboration diversity similarity score for a pair of authors was determined by the absolute difference in their degree centrality values, with lower scores indicating a higher degree of similarity in their collaboration diversity. Conversely, the collaboration diversity level metric reflects the combined diversity of collaborations for the author pair, with higher values indicating greater overall collaboration diversity. This metric is calculated as the sum of their degree centrality values.

**Common publication venues:** Publication venues refer to the conferences and journals where authors have disseminated their work. While authors' disciplines have been identified, this feature provides a more comprehensive representation of an author's field of expertise. It calculates the number of shared journals and conferences in which a pair of authors have published, offering valuable insights into their shared academic interests and potential for future collaboration.

*4.2.2. Structure-based features*

This set of features is derived from the structure of co-authorship networks. A network ($G$) is defined as a tuple ($G=(V,E)$), where ($V$) represents the set of nodes (or vertices), and ($E$) represents the set of edges connecting pairs of nodes within ($V$). An edge between nodes ($x$) and ($y$) is denoted as ($e_{(x,y)}$). The term ($|V|$) indicates the total number of nodes in the network, while ($|E|$) represents the total number of edges. The neighbourhood of a node ($x$), denoted as ($\Gamma_x$), refers to the set of nodes connected to ($x$), and the degree of a node, ($|\Gamma_x|$), is the count of edges linked to it. Our networks are undirected, meaning all edges are bidirectional.

Our structure-based metrics are designed to calculate a score for each pair of authors, reflecting their level of similarity within the network. We employed both *unweighted* and *weighted* local similarity-based link prediction metrics to gain a comprehensive understanding of the factors influencing various co-authorship patterns. For simplicity, we refer to these metrics as unweighted and weighted similarity-based metrics. Unweighted similarity-based metrics evaluate the similarity between two nodes in a network without considering edge weights; for instance, in a co-authorship network, they treat multiple collaborations between two authors as a single collaboration. Conversely, weighted similarity-based metrics take into account the strength of co-authorship links, recognizing that authors with numerous joint collaborations have stronger ties than those with fewer collaborations. In this study, we applied several unweighted similarity-based metrics, including: 1) Common Neighbours (CN), 2) Jaccard Coefficient (JC), 3) Adamic-Adar Index (AA), 4) Preferential Attachment (PA), and 5) Shortest Path (SP). These metrics have been effectively utilized by researchers to predict new co-authorship links (Chuan et al. 2018; Hasan et al. 2006; Pavlov and Ichise 2007; Yu et al. 2014).

**Common Neighbours (CN):** This metric is based on the premise that two nodes with a greater number of shared neighbours are more likely to form a link within a social network. This assumption is supported by the observed correlation between the number of common neighbours and the likelihood of link formation (Newman 2001). The CN metric quantifies the similarity score by counting the shared neighbours, as detailed in Equation (1) (Liben-Nowell and Kleinberg 2007):

$$CN = |\Gamma_x \cap \Gamma_y| \quad (1)$$

where $\Gamma_x$ and $\Gamma_y$ are the sets of authors who have co-authored with author *x* and *y*, respectively, and $|\Gamma_x \cap \Gamma_y|$ is the number of shared co-authors between *x* and *y*. If authors *x* and *y* have many collaborators in common, they are likely to be part of the same research community, which increases the probability of future collaboration.

**Jaccard Coefficient (JC):** Originating from information retrieval systems (Jaccard 1901), the JC metric defines the similarity score as the ratio of the number of shared neighbours to the total set of neighbours of two nodes, as expressed in Equation (2):

$$JC = \frac{|\Gamma_x \cap \Gamma_y|}{|\Gamma_x \cup \Gamma_y|} \quad (2)$$

In the equation, $\Gamma_x \cup \Gamma_y$ is the set of all unique co-authors of *x* and *y* combined, and $|\Gamma_x \cup \Gamma_y|$ is the total number of unique co-authors between *x* and *y*. In co-authorship networks, JC evaluates the similarity between a pair of authors by calculating the ratio of their shared co-authors to the total number of unique co-authors they have combined.

**Adamic-Adar index:** This metric provides a similarity score for a pair of nodes by considering the attributes of their shared neighbours. Unlike the CN metric, the AA index is weighted, penalizing each shared neighbour, (*z*), based on its degree, as described in Equation (3). This adjustment is especially pertinent in scenarios where a node's capacity to allocate resources among its connections decreases as the number of its neighbours grows (Martínez, Berzal, and Cubero 2016).

$$AA = \sum_{z \in \Gamma_x \cap \Gamma_y} \frac{1}{\log |\Gamma_z|} \quad (3)$$

where $\Gamma_z$ is the set of co-authors of author *z*, $|\Gamma_z|$ is the number of co-authors of *z*. The summation runs over all *z* who are common co-authors of *x* and *y*. In co-authorship networks, the AA index assigns higher weights to shared co-authors who have fewer overall co-authors, emphasizing the importance of more exclusive collaborative relationships (Yu et al. 2014).

**Preferential attachment:** The PA metric is based on the principles of power-law distribution and scale-free networks, as introduced by the Barabási-Albert network model (Barabási and Albert 1999). According to this model, the probability of forming a link between two previously unconnected nodes increases with their respective degrees. The PA score, defined in Equation (4), is calculated in our study by multiplying the degrees of the two authors involved.

$$PA = |\Gamma_x||\Gamma_y| \quad (4)$$

**Shortest path:** This metric calculates the minimum number of edges needed to connect pairs of nodes within a network. It operates on the assumption that longer paths, indicated by higher values, decrease the likelihood of forming a future link between two unconnected nodes.

Essentially, authors who are situated further apart in a network are less likely to collaborate than those who are positioned closer together.

In contrast to unweighted similarity-based features, weighted similarity-based metrics offer an additional layer of detail by incorporating collaboration weights into their calculations. The collaboration weight, denoted as ($\omega(a,b)$), quantifies the strength of the link between nodes (*a*) and (*b*), defined by the number of joint papers on which two authors have collaborated (Chuan et al. 2018). For instance, **weighted CN (WCN)** extends the CN measure by incorporating the strength of prior collaborations. Let *x* and *y* be two authors, and $\Gamma_x$ and $\Gamma_y$ represent the sets of their co-authors. The intersection $\Gamma_x \cap \Gamma_y$ is the set of authors who have collaborated with both *x* and *y*. The WCN score is calculated as:

$$WCN = \sum_{z \in \Gamma_x \cap \Gamma_y} \frac{\omega(x,z) + \omega(y,z)}{2} \quad (5)$$

Here, the numerator sums the average collaboration weights between each of the shared co-authors (*z*) and the target authors (*x* and *y*). The WCN score suggests that author pairs with a larger number of shared co-authors, with whom they have published more joint papers, are more likely to engage in future collaborations (Murata and Moriyasu 2007). Similarly, other weighted similarity-based metrics, including the **weighted JC** (**WJC**, Equation (6)), **weighted AA** (**WAA**, Equation (7)), and **weighted PA** (**WPA**, Equation (8)), function under analogous assumptions.

$$WJC = \frac{\sum_{z \in \Gamma_x \cap \Gamma_y} \frac{\omega(x,z) + \omega(y,z)}{2}}{\sum_{u \in \Gamma_x} \omega(x,u) + \sum_{v \in \Gamma_y} \omega(y,v)} \quad (6)$$

$$WAA = \sum_{z \in \Gamma_x \cap \Gamma_y} \frac{\omega(x,z) + \omega(y,z)}{2} \frac{1}{\log(\sum_{t \in \Gamma_x} \omega(z,t))} \quad (7)$$

$$WPA = \sum_{u \in \Gamma_x} \omega(x,u) \sum_{v \in \Gamma_y} \omega(y,v) \quad (8)$$

**Collaboration weight:** We included Collaboration Weight (CW) as a separate metric to measure the strength of collaboration between two authors within a network (Parimi and Caragea 2011). This metric not only considers the number of joint collaborations between authors but also takes into the account the number of authors involved in their shared publications. As outlined in Equation (9), the collaboration weight for an author pair *(x,y)* is adjusted according to the number of authors participating in the joint publications, effectively penalizing larger author groups.

$$CW = \sum_{i=1}^{p} \frac{\sigma_x^i \sigma_y^i}{n_i - 1} \quad (9)$$

In the equation, *p* is the total number of joint publications, $\sigma_x^i$ is equal to 1 if *x* is an author of paper *i*, and it is 0 otherwise. Additionally, $n_i$ represents the number of authors contributing to paper *i*.

### 4.3. Model development

We developed four machine learning classifiers: logistic regression (LR), decision trees (DT), random forest (RF), and extreme gradient boosting (XGBoost). LR serves as a baseline linear model, while DT, RF, and XGBoost are tree-based methods capable of modelling complex feature interactions. We evaluated their performance relative to one another and against a baseline, represented by the prevalence metric. The prevalence metric indicates the proportion of positive samples within the dataset and serves as the baseline for random classification. It reflects the expected accuracy of a classifier that makes predictions entirely at random. We split the dataset into training and testing subsets, allocating 80% of the total data to the training set and the remaining 20% to the testing set. Hyperparameter tuning for the classifiers was conducted using random search. To validate the ML models and determine their optimal hyperparameters, we used repeated stratified 5-fold cross-validation, conducted three times. This approach was chosen to ensure robust performance estimates. The problem was framed as a binary classification task. We assessed the models using several metrics, including precision, recall, F1 score, area under the curve (AUC), and average precision (AP), which calculates the area under the precision-recall curve. AP is particularly recommended for binary classification tasks, especially in addressing imbalanced classification challenges (Davis and Goadrich 2006; Saito and Rehmsmeier 2015). Class imbalance was addressed by random undersampling of the majority class.

### 4.4. Interpretation

We used Shapley Additive Explanations (SHAP) (Lundberg and Lee 2017) to interpret the contribution of each feature to the model's predictions. SHAP values are based on cooperative game theory, where each feature is treated as a "player" contributing to the "payout" (prediction). The SHAP value for a feature represents the average marginal contribution of that feature across all possible feature combinations. Positive SHAP values indicate that the feature increases the likelihood of a link, while negative values suggest the feature decreases it. For example, in predicting new collaborations, a high discipline similarity score might yield a positive SHAP value, increasing the model's confidence that a link will form. SHAP was applied separately to each co-authorship pattern (new, persistent, discontinued) and to the best-performing model for that pattern.

## 5. Results

### 5.1. New co-authorship

#### 5.1.1. Performance evaluation

Table 1 presents the performance of ML models in predicting new collaborations. The average recall values are 0.77 for logistic regression, 0.81 for decision tree, 0.88 for random forest, and 0.87 for the XGBoost model. The random forest classifier achieved the highest average recall, successfully predicting 88% of new co-authorship links on average. The average AUC values for the LR, DT, RF, and XGBoost classifiers are 0.86, 0.84, 0.90, and 0.88, respectively, demonstrating the high performance of the developed models in predicting new co-authorship links. Lower precision values are common in link prediction tasks, as observed in other studies, such as (Chuan et al. 2018). However, these classifiers allow for the adjustment of recall or precision, enabling the predictive model to be tailored to specific needs, albeit at the expense of the other metric.

**Table 1.** Model performance in predicting new collaborations.

| Network | Model | Recall | AUC | AP |
|---|---|---|---|---|
|  | LR | 1.00 | 0.94 | 0.44 |
| 1 | DT | 1.00 | 0.94 | 0.23 |
|  | RF | 1.00 | 0.96 | 0.52 |
| | XGB | 1.00 | 0.85 | 0.30 |
| 2 | LR | 0.68 | 0.91 | 0.76 |
| | DT | 0.90 | 0.88 | 0.55 |
| | RF | 0.87 | 0.92 | 0.73 |
| | XGB | 0.81 | 0.88 | 0.69 |
| 3 | LR | 0.41 | 0.74 | 0.54 |
| | DT | 0.76 | 0.76 | 0.38 |
| | RF | 0.94 | 0.87 | 0.61 |
| | XGB | 0.88 | 0.83 | 0.52 |
| 4 | LR | 0.63 | 0.79 | 0.58 |
| | DT | 0.79 | 0.81 | 0.38 |
| | RF | 0.89 | 0.89 | 0.59 |
| | XGB | 0.74 | 0.88 | 0.52 |
| 5 | LR | 1.00 | 0.96 | 0.49 |
| | DT | 0.95 | 0.92 | 0.25 |
| | RF | 0.95 | 0.97 | 0.74 |
| | XGB | 1.00 | 0.96 | 0.45 |
| 6 | LR | 0.76 | 0.91 | 0.44 |
| | DT | 0.80 | 0.84 | 0.24 |
| | RF | 0.84 | 0.92 | 0.49 |
| | XGB | 0.80 | 0.92 | 0.33 |
| 7 | LR | 0.71 | 0.86 | 0.29 |
| | DT | 0.88 | 0.85 | 0.22 |
| | RF | 0.79 | 0.89 | 0.36 |
| | XGB | 0.83 | 0.89 | 0.31 |
| 8 | LR | 0.85 | 0.80 | 0.22 |
| | DT | 0.60 | 0.80 | 0.17 |
| | RF | 0.75 | 0.85 | 0.42 |
| | XGB | 0.90 | 0.87 | 0.40 |
| 9 | LR | 0.83 | 0.81 | 0.30 |
| | DT | 0.88 | 0.84 | 0.21 |
| | RF | 0.88 | 0.90 | 0.53 |
| | XGB | 0.88 | 0.85 | 0.29 |
| 10 | LR | 0.95 | 0.87 | 0.35 |
| | DT | 0.76 | 0.82 | 0.24 |
| | RF | 1.00 | 0.91 | 0.49 |
| | XGB | 0.95 | 0.89 | 0.41 |
| 11 | LR | 0.94 | 0.91 | 0.38 |
| | DT | 0.78 | 0.85 | 0.25 |
| | RF | 0.94 | 0.89 | 0.42 |
| | XGB | 1.00 | 0.89 | 0.34 |
| 12 | LR | 0.50 | 0.81 | 0.23 |
| | DT | 0.67 | 0.77 | 0.09 |
| | RF | 0.67 | 0.82 | 0.33 |
| | XGB | 0.61 | 0.79 | 0.28 |

**Note:** XGB: XGBoost.

Figure 3 compares the area under the precision-recall curve (AP) of classifiers against the baseline for predicting new co-authorship patterns. All classifiers surpassed the baseline, showcasing their effectiveness in predicting new co-authorship links. The RF classifier demonstrated superior performance across most co-authorship networks, except for the second network, where LR slightly outperformed RF. Notably, the RF model achieved the highest AP value of 0.74 in the fifth co-authorship network, significantly surpassing the baseline value of 0.04.

Following RF, LR and XGBoost were ranked as the second-best algorithms for predicting new co-authorship links.

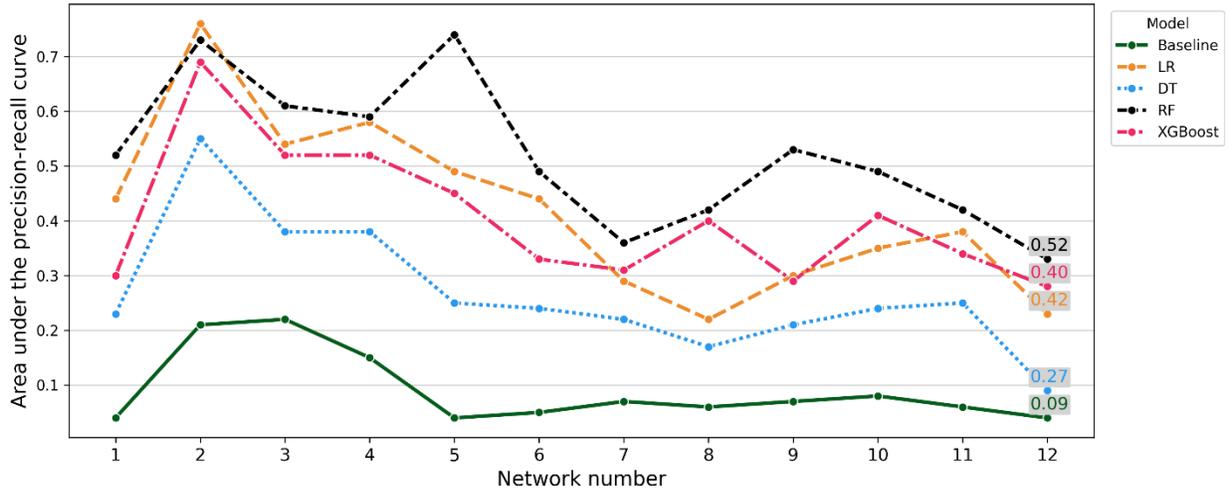

**Figure 3.** Performance evaluation of models in predicting new collaborations based on the area under the precision-recall curve. Values at the end of each line indicate the average performance.

*5.1.2. Driving factors*

We conducted a SHAP analysis (refer to Section 4.4) to pinpoint the key factors driving the prediction of new co-authorship links. Figures 4-a and 4-b illustrate the most influential factors within co-authorship networks 2 and 5, selected due to the models' superior performance in predicting new co-authorship links. In these figures, two colours indicate the impact of the factors: red represents a positive impact, while blue indicates a negative impact on the likelihood of new co-authorship links. The horizontal axis displays the contribution value of each feature, with features ordered in descending order based on their contribution values, placing the most influential factors first. The figures highlight the following features as key predictors: weighted/unweighted Jaccard coefficient, weighted/unweighted Adamic-Adar index, weighted/unweighted common neighbours, weighted/unweighted preferential attachment, discipline similarity score, productivity level of authors, and productivity similarity score.

The features identified above are divided into structure-based and attribute-based factors. Structure-based factors encompass metrics such as the Jaccard Coefficient (JC), Adamic-Adar Index (AA), Common Neighbours (CN), and Preferential Attachment (PA), along with their weighted variants. These metrics are derived from the structure of the co-authorship networks. Research has demonstrated that these structure-based factors play a significant role in the formation of new co-authorship patterns (Liben-Nowell and Kleinberg 2007; Pavlov and Ichise 2007; Yu et al. 2014). Importantly, all structure-based factors, except for both weighted and unweighted preferential attachment metrics, are variations of the CN metric.

**a)**                                              **b)**

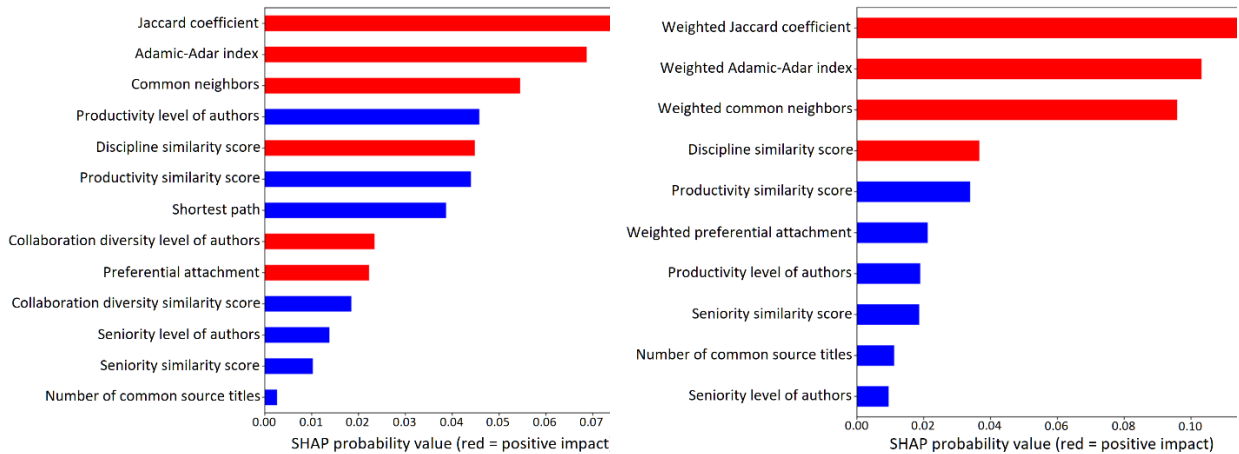

**Figure 4.** SHAP analysis, most important features in predicting new collaboration links: a) Network 2, b) Network 5.

Figures 4-a and 4-b illustrate that these CN-based features positively affect the development of new co-authorship links. The CN metric assesses the number of shared co-authors between two authors, suggesting that authors with more common co-authors are more likely to collaborate in the future. The Jaccard Coefficient, another CN-based metric, is calculated as the ratio of common co-authors to the total number of potential co-authors, with a higher JC indicating a greater likelihood of future collaboration. The Adamic-Adar Index, also CN-based, assigns higher weights to common co-authors with fewer connections themselves. Its positive influence on new co-authorship links suggests that, for two author pairs with an equal number of common co-authors, those whose common co-authors have fewer connections are more likely to form a co-authorship link.

Attribute-based features, including discipline similarity score, productivity level of authors, and productivity similarity score, are also key drivers of new co-authorship patterns. Derived from individual author attributes, the discipline similarity score evaluates the similarity of research fields and disciplines between two authors and positively impacts the likelihood of new co-authorships, as depicted in the figure. This suggests that authors with similar research fields are more inclined to collaborate compared to those from differing disciplines. Furthermore, the productivity similarity score and productivity level of authors are significant attribute-based factors influencing new co-authorship patterns. These metrics use the number of publications to measure an author's productivity. Interestingly, in most networks, these productivity metrics tend to negatively impact the formation of new co-authorship links.

### 5.2. Persistent co-authorship

#### 5.2.1. Performance evaluation

Table 2 presents the performance of various ML models in predicting persistent collaborations. The average recall scores for the classifiers are as follows: LR at 0.75, DT at 0.67, RF at 0.73, and XGBoost at 0.74. The higher precision of these classifiers, compared to those in Table 1, suggests that a substantial portion of the predicted persistent co-authorship links did indeed occur, underscoring their effectiveness in forecasting ongoing collaborations. additionally, the high AUC values reflect the classifiers' strong ability to accurately distinguish between the presence and absence of persistent co-authorship links, highlighting their reliability in predictive tasks.

**Table 2.** Model performance in predicting persistent collaborations.

| Network | Model | Recall | AUC | AP |
|---|---|---|---|---|
| 1 | LR | 0.79 | 0.82 | 0.97 |
| | DT | 0.67 | 0.87 | 0.97 |
| | RF | 0.78 | 0.90 | 0.98 |
| | XGB | 0.76 | 0.88 | 0.97 |
| 2 | LR | 0.81 | 0.69 | 0.95 |
| | DT | 0.68 | 0.79 | 0.95 |
| | RF | 0.73 | 0.84 | 0.98 |
| | XGB | 0.71 | 0.82 | 0.97 |
| 3 | LR | 0.74 | 0.76 | 0.95 |
| | DT | 0.71 | 0.86 | 0.96 |
| | RF | 0.72 | 0.87 | 0.97 |
| | XGB | 0.74 | 0.84 | 0.96 |
| 4 | LR | 0.74 | 0.77 | 0.92 |
| | DT | 0.60 | 0.73 | 0.90 |
| | RF | 0.71 | 0.80 | 0.94 |
| | XGB | 0.68 | 0.78 | 0.93 |
| 5 | LR | 0.73 | 0.74 | 0.92 |
| | DT | 0.64 | 0.78 | 0.92 |
| | RF | 0.73 | 0.84 | 0.96 |
| | XGB | 0.71 | 0.85 | 0.97 |
| 6 | LR | 0.61 | 0.70 | 0.91 |
| | DT | 0.64 | 0.78 | 0.92 |
| | RF | 0.67 | 0.79 | 0.95 |
| | XGB | 0.67 | 0.75 | 0.93 |
| 7 | LR | 0.82 | 0.77 | 0.93 |
| | DT | 0.71 | 0.78 | 0.91 |
| | RF | 0.77 | 0.85 | 0.96 |
| | XGB | 0.81 | 0.82 | 0.95 |
| 8 | LR | 0.78 | 0.79 | 0.92 |
| | DT | 0.70 | 0.76 | 0.91 |
| | RF | 0.69 | 0.81 | 0.94 |
| | XGB | 0.73 | 0.78 | 0.94 |
| 9 | LR | 0.79 | 0.82 | 0.93 |
| | DT | 0.70 | 0.74 | 0.9 |
| | RF | 0.77 | 0.85 | 0.95 |
| | XGB | 0.80 | 0.84 | 0.95 |
| 10 | LR | 0.72 | 0.74 | 0.92 |
| | DT | 0.67 | 0.78 | 0.93 |
| | RF | 0.73 | 0.81 | 0.95 |
| | XGB | 0.76 | 0.85 | 0.96 |
| 11 | LR | 0.67 | 0.79 | 0.93 |
| | DT | 0.68 | 0.79 | 0.93 |
| | RF | 0.71 | 0.86 | 0.97 |
| | XGB | 0.71 | 0.85 | 0.96 |
| 12 | LR | 0.74 | 0.79 | 0.94 |
| | DT | 0.70 | 0.81 | 0.94 |
| | RF | 0.74 | 0.85 | 0.96 |
| | XGB | 0.82 | 0.86 | 0.96 |

Figure 5 illustrates the average precision (AP) performance of the ML models compared to each other and the baseline. All classifiers surpassed the baseline, demonstrating their strong capability in predicting persistent co-authorship links. The classifiers exhibited similar performance, with AP values ranging from 0.89 to 0.97. The average precision for the classifiers was as follows: LR at 0.91, DT at 0.93, RF at 0.94, and XGBoost at 0.93. Similar to the predictions for the new co-authorship pattern, the random forest model generally delivered the highest performance across most co-authorship networks. However, in networks 5 and 10, XGBoost outperformed the RF model.

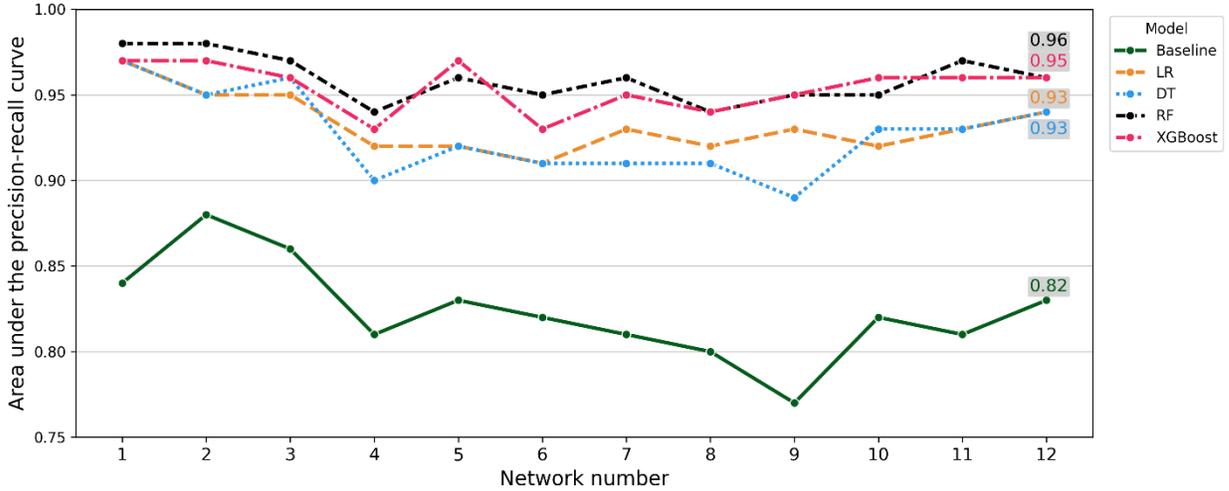

**Figure 5.** Performance evaluation of models in predicting persistent collaborations based on the area under the precision-recall curve. Values at the end of each line indicate the average performance.

### 5.2.2. Driving factors

Figures 6-a and 6-b highlight the most important features in predicting persistent collaboration within co-authorship networks 7 and 11, where the models demonstrated the highest performance. Similar to the new co-authorship pattern analysis (Figure 4), both structure-based and attribute-based metrics emerged as significant predictive features. The key predictive features included authors' productivity, discipline similarity score, seniority level, preferential attachment and its weighted variant, collaboration weight, collaboration diversity similarity score, weighted/unweighted Jaccard coefficient, and weighted/unweighted Adamic-Adar index.

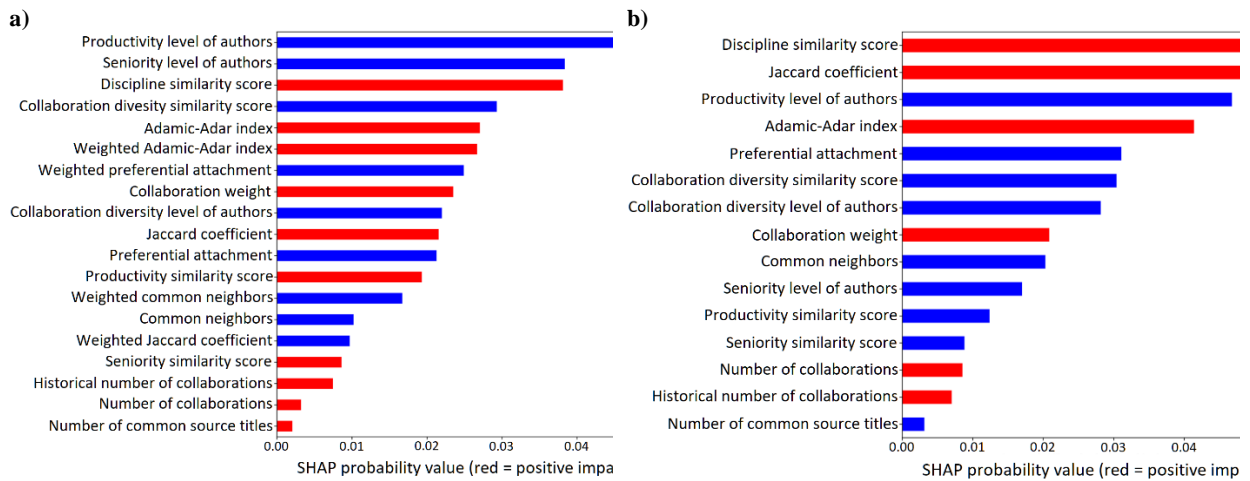

**Figure 5.** SHAP analysis, most important features in predicting persistent collaboration links: a) Network 7, b) Network 11.

The analysis revealed that the productivity level, defined as the total number of publications by an author pair, adversely affects the likelihood of maintaining persistent co-authorship links. Conversely, the discipline similarity score positively influences ongoing co-authorship patterns, indicating that author pairs sharing similar research disciplines are more inclined to collaborate again in the future. Another attribute-based feature, authors' seniority, also negatively impacts persistent co-authorship, suggesting that author pairs with longer career spans are less likely to renew collaborations. Similarly, preferential attachment has a negative effect on the continuity of co-authorship links. This negative influence, along with the impact of authors' productivity levels, could be due to highly productive author pairs being more visible in the research community, thus becoming more accessible to new collaborators and reducing the likelihood of collaborating again with previous partners (Yu et al. 2014).

The collaboration weight, which considers both the number of joint publications by author pairs and the number of authors involved in those publications, positively impacts persistent co-authorship patterns. As shown in Figure 6, this suggests that author pairs with more joint publications involving fewer authors are more likely to collaborate again. Additionally, the collaboration diversity similarity score, a structure-based feature, emerged as a significant predictor of persistent co-authorship patterns. The negative effect of collaboration diversity similarity may imply that authors tend to re-collaborate with those who have similar levels of collaboration diversity. In line with factors driving new co-authorship patterns, both the Jaccard coefficient and Adamic-Adar index positively influence persistent co-authorship patterns, suggesting that author pairs with a greater number of common co-authors are more likely to collaborate again.

### 5.3. Discontinued co-authorship

#### 5.3.1. Performance evaluation

Table 3 showcases the performance of various ML models in predicting the discontinued collaborations. The average recall, which evaluates the model's ability to accurately predict the discontinuation of collaboration between author pairs, was 0.63 for LR, 0.72 for DT, 0.75 for RF, and 0.73 for XGBoost, with RF achieving the highest average recall. In terms of the average AUC, the models performed as follows: LR achieved an AUC of 0.74, DT scored 0.77, RF reached 0.82, and XGBoost attained 0.81. These AUC scores reflect the models' proficiency in distinguishing between the presence and absence of co-authorship links, with RF demonstrating the highest discriminative capability. A higher AUC value indicates better model performance in accurately identifying true positives and negatives, underscoring the effectiveness of RF and XGBoost in predicting discontinued co-authorship patterns.

**Table 3.** Model performance in predicting discontinued collaborations.

| Network | Model | Recall | AUC | AP |
|---|---|---|---|---|
| 1 | LR | 0.71 | 0.67 | 0.40 |
| | DT | 0.93 | 0.89 | 0.48 |
| | RF | 0.79 | 0.86 | 0.64 |
| | XGB | 0.79 | 0.86 | 0.65 |
| 2 | LR | 0.58 | 0.77 | 0.44 |
| | DT | 0.58 | 0.74 | 0.27 |
| | RF | 0.68 | 0.83 | 0.46 |
| | XGB | 0.63 | 0.77 | 0.43 |
| 3 | LR | 0.68 | 0.80 | 0.48 |
| | DT | 0.75 | 0.82 | 0.48 |
|   | RF  | 0.82 | 0.89 | 0.73 |
|   | XGB | 0.82 | 0.89 | 0.61 |
| 4 | LR  | 0.54 | 0.68 | 0.40 |
|   | DT  | 0.63 | 0.66 | 0.30 |
|   | RF  | 0.80 | 0.70 | 0.37 |
|   | XGB | 0.51 | 0.64 | 0.37 |
| 5 | LR  | 0.38 | 0.60 | 0.38 |
|   | DT  | 0.69 | 0.75 | 0.33 |
|   | RF  | 0.67 | 0.79 | 0.50 |
|   | XGB | 0.64 | 0.76 | 0.47 |
| 6 | LR  | 0.66 | 0.68 | 0.41 |
|   | DT  | 0.80 | 0.72 | 0.30 |
|   | RF  | 0.70 | 0.75 | 0.43 |
|   | XGB | 0.68 | 0.77 | 0.46 |
| 7 | LR  | 0.54 | 0.73 | 0.51 |
|   | DT  | 0.75 | 0.78 | 0.41 |
|   | RF  | 0.77 | 0.82 | 0.56 |
|   | XGB | 0.79 | 0.83 | 0.59 |
| 8 | LR  | 0.78 | 0.86 | 0.68 |
|   | DT  | 0.76 | 0.79 | 0.42 |
|   | RF  | 0.84 | 0.87 | 0.64 |
|   | XGB | 0.88 | 0.86 | 0.60 |
| 9 | LR  | 0.55 | 0.76 | 0.62 |
|   | DT  | 0.67 | 0.76 | 0.45 |
|   | RF  | 0.65 | 0.84 | 0.69 |
|   | XGB | 0.69 | 0.79 | 0.61 |
| 10| LR  | 0.62 | 0.77 | 0.49 |
|   | DT  | 0.67 | 0.80 | 0.46 |
|   | RF  | 0.73 | 0.84 | 0.61 |
|   | XGB | 0.81 | 0.85 | 0.57 |
| 11| LR  | 0.80 | 0.82 | 0.54 |
|   | DT  | 0.78 | 0.82 | 0.48 |
|   | RF  | 0.81 | 0.86 | 0.61 |
|   | XGB | 0.85 | 0.88 | 0.64 |
| 12| LR  | 0.72 | 0.79 | 0.46 |
|   | DT  | 0.58 | 0.75 | 0.36 |
|   | RF  | 0.70 | 0.81 | 0.55 |
|   | XGB | 0.66 | 0.79 | 0.50 |

Figure 7 illustrates the average precision performance of the classifiers compared to the baseline, demonstrating that all classifiers surpassed the baseline in predicting discontinued co-authorship links. Among them, RF and XGBoost delivered the best performance.

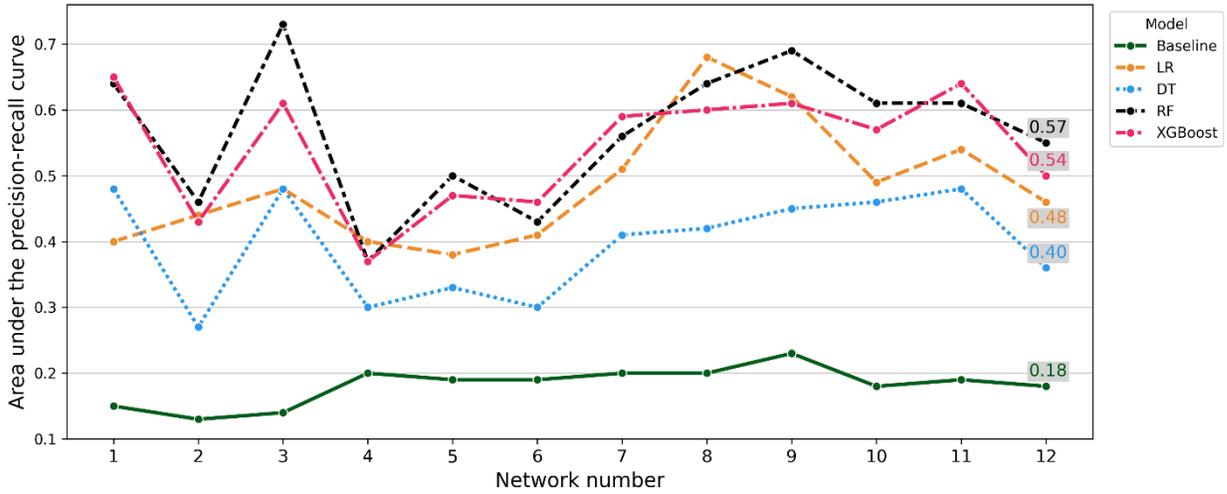

**Figure 7.** Performance evaluation of models in predicting discontinued collaborations based on the area under the precision-recall curve. Values at the end of each line indicate the average performance.

### 5.3.2. Driving factors

Figure 8 highlights the key features influencing discontinued collaborations. Unlike the trends observed in persistent co-authorship patterns, the productivity level of authors has a positive effect on the likelihood of collaborations ending. Additionally, the seniority level of authors, calculated as the combined career age of author pairs, emerges as a significant factor, also positively impacting discontinued co-authorships. This suggests that author pairs with longer collective career spans are more prone to ending their collaborations compared to those who are earlier in their careers. On the other hand, collaboration weight negatively affects the likelihood of co-authorships ending. This implies that author pairs with a larger number of joint publications involving fewer contributors are less likely to terminate their collaboration. This tendency may be attributed to the strong collaborative bonds these author pairs have previously formed, making them more inclined to continue working together.

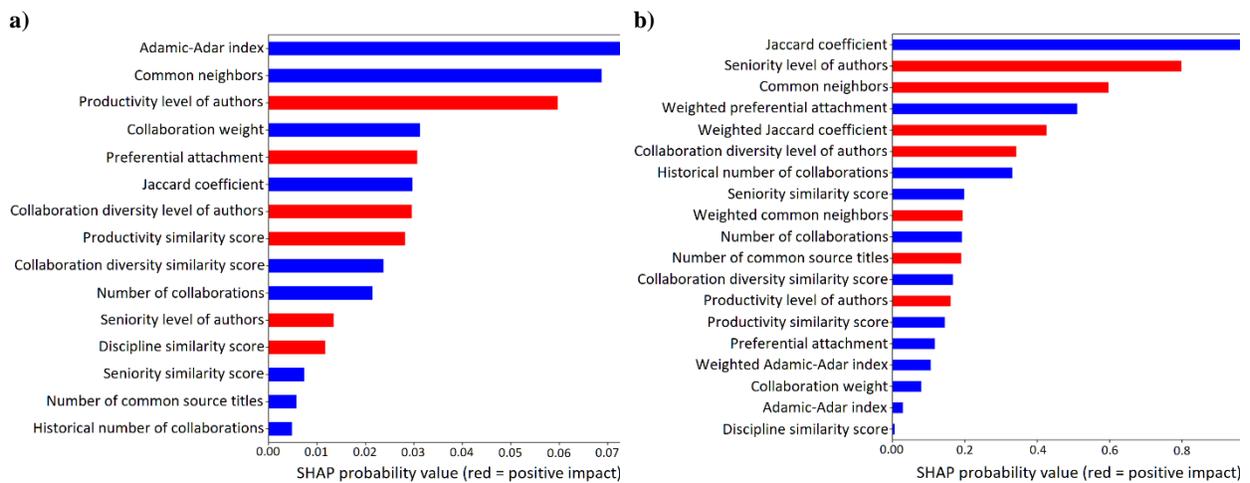

**Figure 8.** SHAP analysis, most important features in predicting discontinued collaboration links: a) Network 3, b) Network 8.

Structure-based metrics, including unweighted and weighted Jaccard coefficient, Adamic-Adar index, common neighbours, preferential attachment, and weighted preferential attachment, influence discontinued co-authorship patterns. However, their impact is not consistently positive or negative, suggesting that a higher number of common co-authors does not necessarily correlate with a lower likelihood of author pairs ceasing collaboration in the future. This may indicate that the relationship between these metrics and co-authorship discontinuation is complex and context-dependent.

## 6. Conclusion

This study set out to predict various co-authorship links and uncover key factors influencing different co-authorship patterns among researchers, using "AI-driven cancer research" as a case study to illustrate the effectiveness of our approach. By extracting both structure-based and attribute-based metrics from co-authorship networks and individual authors, we used these features as inputs for machine learning classifiers to predict new, persistent, and discontinued co-authorship patterns. The classifiers exhibited strong predictive performance, highlighting the value of integrating both types of metrics for accurate co-authorship link prediction.

Our analysis revealed that attribute-based metrics, such as authors' productivity levels and seniority, play a significant role in shaping these patterns. While high productivity negatively affects persistent co-authorship, it positively correlates with discontinuation, suggesting that prolific authors may seek new collaborations. Similarly, seniority positively influences discontinued co-authorships, indicating that authors with longer careers might be more inclined to diversify their collaborations. Additionally, the discipline similarity score emerged as a crucial factor, positively influencing new and persistent patterns while negatively affecting discontinued patterns. This suggests that authors working in similar fields are more likely to initiate or repeat collaborations and are less likely to end them.

Structure-based metrics, such as the Jaccard coefficient, Adamic-Adar index, and preferential attachment, also emerged as key influencers, though their effects varied across different co-authorship patterns. This complexity underscores the nuanced nature of co-authorship dynamics, where sharing common co-authors does not necessarily predict ongoing or future collaboration.

Our findings emphasize the importance of understanding both attribute-based and structure-based factors when analyzing author collaboration networks. The robust performance and interpretability of the classifiers can help researchers worldwide identify potential collaborators and assist research organizations in forming cohesive, effective research teams. This study not only enhances the understanding of co-authorship dynamics but also supports the idea that co-authorship arises from social networks, offering valuable insights into building strong academic collaborations.

## 7. Limitations and future work

This study encountered several limitations, mainly stemming from the data source used. While Scopus and PubMed are recognized as comprehensive resources in the medical field, as noted by Falagas et al. (Falagas et al. 2008), incorporating additional sources could provide a more comprehensive perspective. Furthermore, excluding non-English publications might have restricted the diversity of the dataset.

It is important to note that co-authorship does not always imply collaboration. Authors may collaborate without resulting in co-authorship, and not all co-authored papers result from genuine collaboration (Ebadi and Schiffauerova 2015). In our study, structure-based metrics were derived from "homogeneous" co-authorship networks, which consist of a single type of node (authors) and edge (co-authorship links). These metrics were employed as input for machine learning classifiers and the SHAP approach to identify collaboration factors. Future research could benefit from constructing "heterogeneous" co-authorship networks that incorporate various node types (e.g., authors, institutes, papers, fields, venues) and edge types (e.g., author roles within papers, affiliations, publication venues, fields, citations). These complex networks would allow for the extraction of more nuanced features, enabling advanced techniques for co-authorship link prediction and offering deeper insights into the factors driving co-authorship, as suggested by Hu et al. (Hu et al. 2020). Such developments could significantly enhance our understanding of academic collaboration dynamics and improve predictive models.

Lastly, future studies could enhance the current framework by explicitly classifying author affiliations into categories such as academic, private, governmental, and non-profit, enabling the examination of sector-specific collaboration patterns and revealing potential structural differences between private–public partnerships and purely academic collaborations in AI-driven cancer research.